\newif\ifmnras
\mnrastrue
\ifmnras
\documentclass[useAMS,usenatbib,usedcolumn]{mnras}
\else
\documentclass[apj,numberedappendix]{emulateapj}
\fi
\usepackage{graphics,epsf}
\usepackage{amsmath}                
\usepackage{amsfonts}               
\usepackage{amssymb}                
\usepackage{epsfig}                 
\usepackage{graphicx}
\usepackage{float}

\newcommand{\km}{{~\rm km}}
\newcommand{\s}{{~\rm s}}

\newcommand{\yr}{{~\rm yr}}

\newcommand{\ourlongtitle}{{Type~IIb supernova progenitors by fatal common envelope evolution}}
\newcommand{\ourshorttitle}{{SNe~IIb by fatal CEE}}

\ifmnras
\else

\newcommand{\na}{{~\rm New Astronomy}}

\fi

\ifmnras
\title[\ourshorttitle]{\ourlongtitle}
\author[N. Lohev, E. Sabach, A. Gilkis, N. Soker]{Noam Lohev$^{1}$\thanks{Contact e-mail: \href{nlohev@campus.technion.ac.il}{nlohev@campus.technion.ac.il}}, Efrat Sabach$^{1}$\thanks{Contact e-mail: \href{efrats@physics.technion.ac.il}{efrats@physics.technion.ac.il}}, Avishai Gilkis$^{2}$\thanks{Contact e-mail: \href{agilkis@ast.cam.ac.uk}{agilkis@ast.cam.ac.uk}}, Noam Soker$^{1,3}$\thanks{Contact e-mail: \href{soker@physics.technion.ac.il}{soker@physics.technion.ac.il}}
\\
$^{1}$ Department of Physics, Technion -- Israel Institute of Technology, Haifa 3200003, Israel \\
$^{2}$ Institute of Astronomy, University of Cambridge, Madingley Road, Cambridge, CB3 0HA, UK  \\
$^{3}$ Guangdong Technion Israel Institute of Technology, Shantou, Guangdong Province 515069, China
}
\fi
 
\begin{document}

\ifmnras
\label{firstpage}
\pagerange{\pageref{firstpage}--\pageref{lastpage}} \pubyear{2019}
\maketitle
\else
\title{\ourlongtitle}
\author{Noam Lohev\altaffilmark{1}, Efrat Sabach\altaffilmark{1}, Avishai Gilkis\altaffilmark{2} \& Noam Soker\altaffilmark{1,3}}
\altaffiltext{1}{Department of Physics, Technion -- Israel Institute of Technology, Haifa
3200003, Israel; nlohev@campus.technion.ac.il; efrats@physics.technion.ac.il; agilkis@ast.cam.ac.uk; soker@physics.technion.ac.il}
\altaffiltext{2}{Institute of Astronomy, University of Cambridge, Madingley Road, Cambridge, CB3 0HA, UK; agilkis@ast.cam.ac.uk}
\altaffiltext{3}{Guangdong Technion Israel Institute of Technology, Shantou, Guangdong Province 515069, China}
\fi

\begin{abstract}
From stellar evolution simulations (using \textsc{mesa}) we conclude that the fatal common envelope evolution (CEE) channel for the formation of Type~IIb core collapse supernova (SN~IIb) progenitors can indeed account for some SNe~IIb. In the fatal CEE channel for SNe~IIb a low mass main sequence secondary star inspirals inside the giant envelope of the massive primary star and removes most of the giant envelope before it merges with the giant core. The key ingredient of the scenario studied here is that the tidally destroyed secondary star forms a new giant envelope. The mass-loss process in a wind during the evolution from the merger process until core collapse, i.e., until the explosion, leaves little hydrogen mass at explosion as inferred from observations of SNe~IIb. In the case of a massive primary star with a zero age main sequence mass of $M_{\rm ZAMS} = 16 M_\odot$ that during its giant phase swallows a main sequence star of mass $M_2=0.5 M_\odot$, we find at explosion a hydrogen mass of  $M_{\rm H} \simeq 0.02$--$0.09 M_\odot$, depending on the rotation we assume. We find similar values for $M_{\rm ZAMS} = 12 M_\odot$. 

\ifmnras
\else
\fi
\end{abstract}

\ifmnras
\fi

\section{INTRODUCTION}
\label{sec:intro}

Massive stars explode as core-collapse supernovae (CCSNe) with varying amounts of hydrogen in their envelope. Those that have envelopes with a small amount of hydrogen show strong hydrogen lines at early times and very weak hydrogen lines, or even none, at later times. These are classified as Type~IIb supernovae (SN~IIb). \cite{Sravanetal2019} take the mass of the hydrogen-rich envelope of the progenitor at the onset of explosion to be $0.01 M_\odot \le M_{\rm H} \le 1 M_\odot$ (also \citealt{Yoonetal2017}), while others take somewhat narrower ranges of $M_{\rm H} \simeq 0.03$--$0.5 M_\odot$ (e.g., \citealt{Woosleyetal1994, Meynetetal2015}). 

{{{{ There are basically two types of SNe~IIb progenitors, compact progenitors and extended progenitors, i.e., red supergiants \citep{ChevalierSoderberg 2010}. \cite{Yoonetal2017} study the formation of SNe~IIb by Roche lobe overflow (RLOF) mass transfer while considering three groups of SNe~IIb, namely, blue progenitors, yellow supergiants, and red supergiants. The more compact blue progenitors and yellow supergiants have little hydrogen mass at explosion, $M_{\rm H} \la 0.15 M_\odot$ and mostly with low metallicity, while the red supergiants have a hydrogen mass of $M_{\rm H} \ga 0.15M_\odot$ at explosion. Compact progenitors of SNe~IIb result from both stable and unstable mass transfer, and probably are the majority of SNe~IIb. \cite{Yoonetal2017} further comment that in the case of RLOF models for SNe~IIb there is a post-RLOF wind that removes more hydrogen, even to the point of forming a SN~Ib (see also \citealt{Gilkisetal2019}). Namely, there is continuous variation of hydrogen mass from compact SNe~IIb progenitor to SNe~Ib, with higher metallicity populations removing more mass and hence having a higher ratio of SNe~Ib to SNe~IIb. \cite{Yoonetal2017} take a ratio between the initial secondary and primary mass of $q=0.9$ in all their models. Binary systems with lower mass ratios have problems to account for SNe~IIb, in particular those with extended envelopes having $M_{\rm H} \ga 0.15M_\odot$ (\citealt{Podsiadlowskietal1992}). One of the advantages of the model we study here is that it does not require a mass ratio close to one to form an extended progenitor of SNe~IIb. }}}}
    
Studies estimate the fraction of SNe~IIb out of all CCSNe to be $f_{\rm IIb} \simeq 11 \%$ (e.g., \citealt{Smithetal2011, Shivversetal2017, Grauretal2017}). In a recent population synthesis study \cite{Sravanetal2019} take $f_{\rm IIb, L} \simeq 20 \%$ in low-metallicity stellar populations and $f_{\rm IIb,H} \simeq 10-12 \%$ in high-metallicity stellar populations.  \cite{Sravanetal2019} study both single and binary stellar evolutionary routes and find that the two contribute about equally to the population of SNe~IIb progenitors. However, the combined contribution of single and binary systems in their calculations is about a factor of three or more lower than the observed rate of SNe~IIb (also \citealt{Sravan2016}). {{{{ This tension might be alleviated if the mass-loss rate after RLOF is lower than usually assumed \citep{Gilkisetal2019}. }}}}

In the present study we focus on binary stellar evolution. Observations support the notion that, at least a large fraction of, SNe~IIb come from binary stellar interactions. \cite{Kilpatricketal2017} could fit a binary model for the progenitor of SN~2016gkg where the mass of the primary star at explosion was $M_{1,f}=5.2 M_\odot$, after it lost most of its mass in a binary interaction. The initial masses were $M_{1,i}=15M_\odot$ and $M_{2,i}=1.5 M_\odot$ and the initial period was $P_i= 1000~$days. Other examples include \cite{Benvenutoetal2013} who fit a binary progenitor for the SN~IIb 2011dh, and the recent paper by \cite{Nakaokaetal2019} who fit a binary model to the SN~IIb 2017czd. \cite{Podsiadlowskietal1993} suggested that the progenitor of SN~IIb 1993J was a binary system. Later \cite{Alderingetal1994} supported this suggestion from the photometry of this SN~IIb. \cite{Foxetal2014} use the flattened circumstellar matter around SN~1993J \citep{Mathesonetal2000} to argue for a stellar binary progenitor.

{{{{ When it comes to binary scenarios for SNe~IIb the most studied scenarios involve mass transfer, mainly by RLOF, that reduces the hydrogen envelope to a very low mass. \cite{Jossetal1988} already discussed such a process as a possible way to form a SN~II that explodes as a blue giant. \cite{Nomotoetal1993} and \cite{Podsiadlowskietal1993} suggest mass transfer in a binary system to account for the low hydrogen mass of SN~IIb 1993J. \cite{Podsiadlowskietal1992} estimate that mass transfer leads to SNe~IIb in only about one percent of CCSNe. Overall, SNe~IIb that belong to the compact type of progenitors, that have $M_{\rm H} \la 0.15 M_\odot$, probably account for the majority of SNe~IIb. }}}}
  
\cite{Claeys2011} study a mass transfer scenario by expanding the work of \cite{StancliffeEldridge2009}. They find that the binary evolution they consider can explain only about five per cent of all SNe~IIb, but this fraction increases if the companion accretes only a small fraction of the mass that is lost from the binary system (see also \citealt{OuchiMaeda2017}), and if the mass outflow carries relatively low angular momentum. 

\cite{Soker2017} takes the above properties, of low accretion mass by the companion, of low angular momentum of the ejected mass, and of a flattened mass loss, as supporting evidence for the grazing envelope evolution (GEE) scenario for the progenitors of some SNe~IIb. In the GEE the companion grazes the giant envelope and launches jets. The jets remove mass from the envelope. Some of these properties, such as mass ejection and a flattened outflow, are shared with post-asymptotic giant branch intermediate binaries (post-AGBIBs; e.g., \citealt{Kastneretal2010, VanWinckel2017}), where observations in some systems find the main sequence (MS) companion that closely orbits the post-AGB star to launch jets (e.g., \citealt{Wittetal2009, Gorlovaetal2012, Thomasetal2013, Gorlovaetal2015, VanWinckel2017b}). The GEE is an additional mass transfer scenario, different from the RLOF scenario, and hence expands the binary parameter space that can lead to SNe~IIb. 

We suggest that binary stellar evolution explains most, or even all, SNe~IIb. For that to be the case we add two more evolutionary channels in addition to the RLOF scenario that most studies, e.g., \cite{Sravanetal2019}, consider.
The first evolutionary channel, as we described above, is the GEE as proposed by \cite{Soker2017} and further studied by \cite{Naimanetal2019} in a recent paper. In the second evolutionary channel a MS companion ejects all, or most, of the original hydrogen-rich envelope of the SN~IIb progenitor, but then it suffers a fatal merger with the core of the giant. The companion is destroyed on the core and becomes the new low-mass hydrogen-rich envelope of the massive star. This fatal common envelope evolution (CEE) scenario for SNe~IIb is the subject of the present study. 
    
There are different types of mergers of a core of a giant star with the more compact secondary star, sometimes resulting in a fatal destruction of the companion on the core (for a review of some fatal CEE evolutionary channels see \citealt{Soker2019FatalCEE}). The companion itself can be a substellar object, like a brown dwarf (e.g., \citealt{HarpazSoker1994, SiessLivio1999}), a MS star, e.g., as in some scenarios for the progenitor of SN~1987A (e.g., \citealt{ChevalierSoker1989, Podsiadlowskietal1990, MenonHeger2017, Urushibataetal2018, Menonetal2019}) or, as in a scenario for unusual nucleosynthesis \citep{IvanovaPodsiadlowski2002}, a white dwarf (e.g., \citealt{IlkovSoker2012} for the core degenerate scenario of Type~Ia supernovae), or the companion can be the result of a supernova, i.e., a neutron star or a black hole \citep{Chevalier2012}. The merger of a neutron star with the core can be a very violent event and lead to a supernova-like event \citep{Chevalier2012} that is termed a common envelope jets supernova \citep{SokerGilkis2018}. The merger of the two cores of two giants can lead to a bright transient event with massive circumstellar matter \citep{Segevetal2019}.

In the present paper we consider the formation of a SN~IIb progenitor from a fatal CEE. We suggest that it can account for Cassiopeia~A. From the light echo \cite{Restetal2011} deduced that the supernova Cassiopeia~A was a SN~IIb. However, at explosion the star was probably a single star \citep{Kochanek2018,Kerzendorfetal2019}. \cite{Nomotoetal1995} suggested a merger scenario for SNe~IIb. In their model the inspiral of the companion inside the envelope removes a large mass from the envelope. Similarly, \cite{Youngetal2006} considered a fatal CEE for the progenitor of Cassiopeia~A.  We differ by that we discuss a scenario where first the entire (or almost all of the) envelope of the giant is removed, and then the secondary star is destroyed and supplies the envelope that will be present at explosion.
In section \ref{sec:Numerics} we describe the numerical evolutionary code, in section \ref{sec:massremoval} we describe the structure of the giant stars before mass removal, and in section \ref{sec:merger} we study the evolution of the core-companion merger product until core collapse. We present our summary in section \ref{sec:summary}. 

\section{Numerical setup}
\label{sec:Numerics}

We mimic the binary interaction leading to a fatal CEE step by step for a massive star and a low mass companion, both with an initial metallicity of $Z=0.02$. We evolve a single star using the \textsc{mesa} code (Modules for Experiments in Stellar Astrophysics, version 10398; \citealt{Paxtonetal2011, Paxtonetal2013, Paxtonetal2015, Paxtonetal2018}) and break the evolution of the primary star to 5 steps. 
(1) pre-main sequence (pre-MS) to zero age MS (ZAMS);
(2) ZAMS to the point where binary interaction is most likely to take place, when the primary reaches the giant phase;
(3) mimicking the secondary inspiral: mass loss of the primary envelope;
(4) mimicking the merger: mass accretion of the secondary star (mass addition phase);
(5) post-merger evolution until collapse.

We note the following caveats in our mass transfer treatment. First, though we do not know how long the mass removal process lasts (but probably no more than tens or maybe hundreds of years) we here take a period of $\approx 2000 \yr$ that assures numerical convergence. Second, while mass transfer is a three-dimensional (3D) process involving angular momentum, we here simulate the interaction using a 1D code for the mass loss and mass accretion phases which we set.

To assess the role of rotation in our scenario we examine two cases. We simulate one case without rotation along the entire evolution, and one case with rotation. In the case with rotation the ZAMS equatorial velocity is $v_{\rm rot,i}=100 \km \s^{-1}$. Spin-down is included by way of angular momentum being carried away by the stellar wind, with the mass-loss rate following \cite{deJager1988} when $T_\mathrm{eff} \le 10\,000\,\mathrm{K}$, according to \cite{Vink2001} when $T_\mathrm{eff} \ge 11\,000\,\mathrm{K}$, and interpolating in between. We do not include the spin-up of the giant envelope when the secondary star spirals in. The material accreted in the merger-mimicking phase has a specific angular momentum of $j_\mathrm{acc}=4\times 10^{18} \, \mathrm{cm}^2\,\mathrm{s}^{-1}$, resulting in an equatorial rotation velocity of about $7 \%$ of the breakup velocity at the end of the mass addition phase for the model with a primary star of $M_{\rm ZAMS} = 16 M_\odot$ model, and about $5 \%$ for the model with a primary star of $M_{\rm ZAMS} = 12 M_\odot$.

\section{Mass removal}
\label{sec:massremoval}

\begin{figure}
{\includegraphics[scale=0.46]{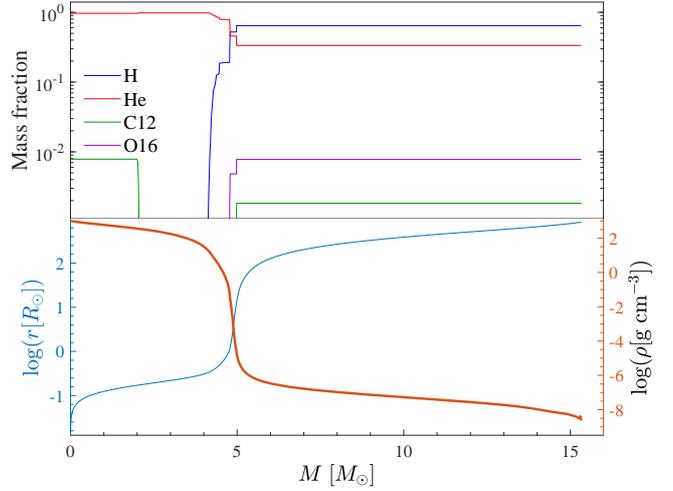}}
\caption{The profiles of some quantities of the $M_{\rm ZAMS} = 16 M_\odot$ stellar model with rotation, when we start to remove the envelope. In the upper panel we present the composition of the main elements and in the lower panel we present the radius (blue-thin line) and  density profile (orange-thick line) as a function of mass.}
\label{fig:initial16}
\end{figure}

We first describe the evolution of a star with a ZAMS mass of $M_{\rm ZAMS}=16 M_\odot$ and with rotation ($v_{\rm rot,i}=100 \km \s^{-1}$; see section \ref{sec:Numerics}). We assume that after the star undergoes its large expansion it swallows a secondary star of mass $M_2 \approx {\rm several} \times 0.1 M_\odot$. In particular, we here take $M_2=0.5M_\odot$ to enter the giant star envelope when the giant radius reaches a maximum value of $R_1=848 R_\odot$. In Fig. \ref{fig:initial16} we present the primary stellar model at that stage, when we assume that the secondary star enters the giant envelope and spirals in, and hence we start to rapidly remove envelope mass.

\begin{figure}
{\includegraphics[scale=0.44]{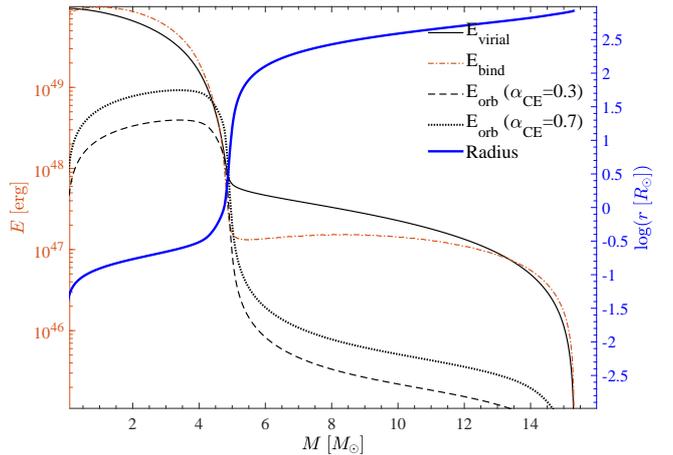}}
\caption{Some quantities relevant to envelope mass removal at the onset of the CEE for the $M_{\rm ZAMS} =16 M_\odot$ stellar model with rotation (Fig. \ref{fig:initial16}). In the thick solid blue line we present the radius as a function of mass, and in the dashed-dotted orange line we present the binding energy of mass residing above radius $r$. We also plot the gravitational energy that the inspiralling secondary star of mass $M_2=0.5 M_\odot$ deposits to the envelope for two values of the common envelope parameter $\alpha_{\rm CE}$ as indicated in the inset.}
\label{fig:binding}
\end{figure}

The core tidally destroys the secondary star at a radius $r_t$ where the average density of the secondary star is about equal to the average density of the primary star inner to that radius. We find this radius to be about $r_t \simeq 1.5 R_\odot$. The mass coordinate is $M(1.5 R_\odot)=4.8 M_\odot$. 
We first need to ensure that the companion can in principle remove the envelope above that radius. In Fig. \ref{fig:binding} we present the binding energy of the envelope above radius $r$ as a function of the corresponding mass coordinate as computed by the integration of the total energy (thermal plus gravitational) from the surface inwards to radius $r$ (dashed-dotted orange line). For comparison we also present the energy $E_{\rm virial}(r)= -0.5 E_G(r)$, where $E_G(r)$ is the gravitational energy of the envelope mass above that radius (black solid line).

We also plot the gravitational energy that the companion of mass $M_2=0.5 M_\odot$ releases to envelope removal as it inspirals to a radius $r$, $E_2=\alpha_{\rm CE} GM(r)M_2/(2r)$, where $\alpha_{\rm CE}$ is the common envelope parameter and $M(r)$ is the mass of the primary star (giant) inner  to radius $r$. From Fig. \ref{fig:binding} we learn that the companion can remove the envelope gas residing above radius $r_t \simeq 1.5 R_\odot$ for $\alpha_{\rm CE} M_2 \ga 0.2 M_\odot$. We therefore remove the entire envelope mass above the mass coordinate $M=4.8 M_\odot$. 
  
The orbital period of the companion on the surface of the giant star is about two years. We expect the inspiral of the secondary star down to the core to last for approximately tens of years. Due to numerical reasons though, we remove the envelope of the giant star, of $10.5 M_\odot$, within a time of about $ 2000 \yr $ at a constant rate. 

At the end of the envelope removal phase, before we add the mass of the secondary star, the giant in our simulation with rotation has a hydrogen mass of $0.11 M_\odot$, while in the simulation without rotation the hydrogen mass is $0.10 M_\odot$.

 
\section{Merger to explosion}
\label{sec:merger}

After most of the hydrogen-rich envelope is removed (section \ref{sec:massremoval}) we accrete the companion to the (almost) bare core, namely, we add a mass of $M_{\rm add} = M_2=0.5 M_\odot$ and the same composition as the initial composition of the primary star. Including the leftover hydrogen, the merger product has a new hydrogen-rich envelope with a hydrogen mass of $\simeq 0.46 M_\odot$. 

The destruction of the companion occurs over several times the dynamical time of the core-companion binary system, amounting to about few days and less. But, again, due to strong numerical limitations we add the mass at a slow rate over a {{{{ time of $t_{\rm add} = 50 \yr$. Although much longer than the dynamical time, it is shorter than the thermal time scale of the new envelope (about $10^3 \yr$) and much shorter than the rest of the evolutionary time until explosion (about $10^6 \yr$). Adding the mass over a time of $ 500 \yr $ gives basically the same evolution. }}}}

In the simulation with rotation that we present here at the end of the mass addition phase the equatorial rotation velocity is about $7 \%$ of the break up rotation velocity. As with young stellar objects that accrete mass, the merger product can lose a large amount of angular momentum by magnetic activity that blows off a small amount of mass. Namely, the merger product might have a slow rotation. We simulated therefore a case with no rotation, and also one case with a faster rotation and one case with a slower rotation (that we do not present here). 

\begin{figure}
{\includegraphics[scale=0.46]{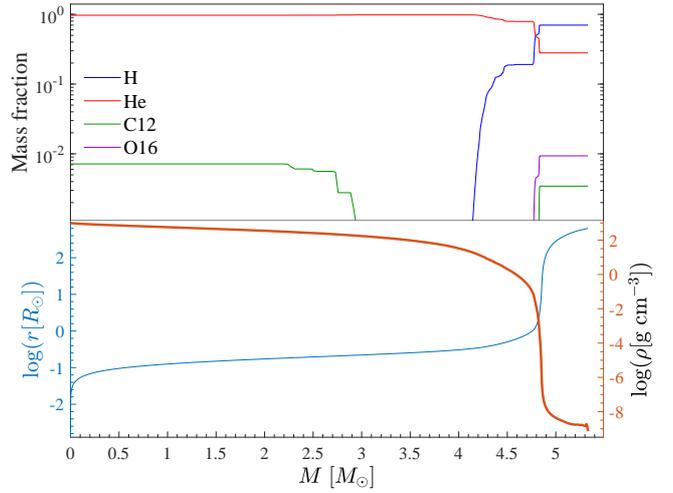}}
\caption{The profiles of some quantities in the $M_{\rm ZAMS} = 16 M_\odot$ stellar model with rotation when we just finish adding the entire secondary star of mass $M_2=0.5 M_\odot$ on to the almost bare core of the giant. Lines are as in Fig. \ref{fig:initial16}. }
\label{fig:AfterAdd16}
\end{figure}

In Fig. \ref{fig:AfterAdd16} we present the structure of the star (merger product) in the rotating simulation at the end of the mass addition phase.   

We now follow the star until core collapse. The final mass of the envelope, in particular the final hydrogen mass, strongly depends on the mass-loss rate. Because the envelope spins-down with mass loss, the final hydrogen mass at explosion only weakly depends on the initial rotation (numerically this is true as long as there is rotation; see below). Therefore in this study we present in detail only the simulation which starts with rotation on the MS and after merger reaches a rotation of 7 per cent of the break up rotation. 

We find that the hydrogen mass at core collapse is $M_{\rm H,exp,rot}=0.085 M_\odot$ for the evolution with rotation at $7\%$ break up velocity after mass addition. For $0.5\%$ and $30\%$ break up velocity after mass addition the leftover hydrogen mass is the same. For the simulation with no rotation at all we find $M_{\rm H,exp}=0.024 M_\odot$, which we consider a special (numerically) case. This results from the envelope mass being lower by the same factor. These hydrogen masses fall in the range of SNe~IIb (section \ref{sec:intro}). 

\begin{figure}
{\includegraphics[scale=0.46]{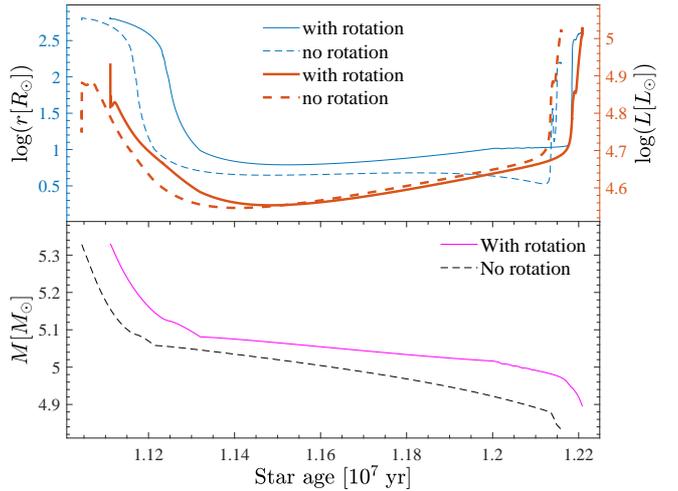}}
\caption{Evolution of the merger products of the $M_{\rm ZAMS} = 16 M_\odot$ star (after mass removal) with an $M_2=0.5 M_\odot$ companion. We show the results for the models with (solid lines) and without (dashed lines) rotation, from the end of the numerical mass addition phase to core collapse. In the upper panel we show the radius (blue-thin lines) and luminosity (orange-thick lines) and in the lower panel we show the total mass.}
\label{fig:RadiusEvolution16}
\end{figure}

In Fig. \ref{fig:RadiusEvolution16} we present the radius, luminosity, and mass of the two models we describe here, with and without rotation, from the end of the mass addition phase to core collapse. Note that the two models do not start this phase at the same time due to different evolution times until mass removal (the model without rotation evolves faster). Besides the age, the differences between the two models are very small.    

\begin{figure}
{\includegraphics[scale=0.46]{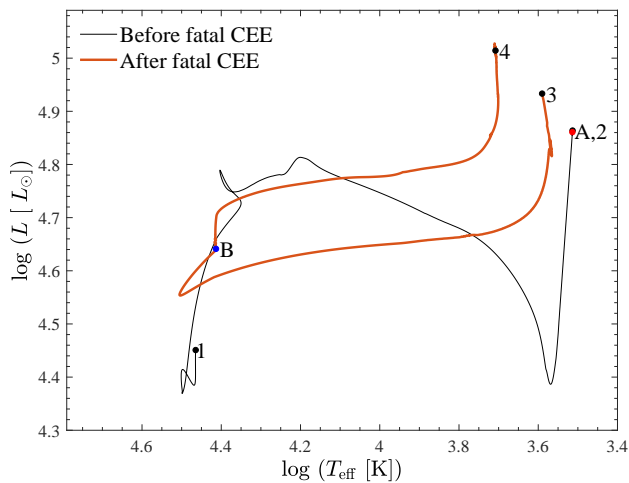}}
\caption{The evolution on the HRD of the $M_{\rm ZAMS} = 16 M_\odot$ star with rotation before and after the CEE. The black line shows the evolution before mass removal starts, from the ZAMS (point 1) to the beginning of the mass removal phase (point 2), and the orange line shows the evolution from the end of the mass addition phase (point 3) to core collapse (point 4). {{{{ Points A and B mark the beginning and end of core helium burning, respectively. }}}} }
\label{fig:HRAdd16}
\end{figure}

In Fig. \ref{fig:HRAdd16} we present the evolution of the $M_{\rm ZAMS} = 16 M_\odot$ star on the Hertzsprung-Russell diagram (HRD). Because our phases of mass removal and then mass addition are set by numerical limitations we do not show them. We show two phases of evolution, from ZAMS to the beginning of mass removal (black line, points 1 to 2), namely the onset of the CEE, and from the end of mass addition to core collapse (orange line, points 3 to 4). {{{{ Points A and B mark the beginning and end of core helium burning, respectively. }}}} 

Overall, the evolution on the HRD is similar, but not identical, to that of stars that start on the ZAMS with a mass of $\simeq 10-20 M_\odot$  (e.g., fig. 2 of \citealt{Georgyetal2013}). Before mass removal our model is not different from other models (e.g., the $9 M_\odot$ model of \citealt{Georgyetal2013}). After we remove most of the envelope and add only $0.5 M_\odot$, the star becomes somewhat bluer than single stars that suffer no rapid mass loss. As the star here keeps some hydrogen-rich envelope, at its final evolutionary stage before core collapse it expands and becomes redder (e.g., the $20 M_\odot$ model of \citealt{Georgyetal2013}).

\begin{figure}
{\includegraphics[scale=0.46]{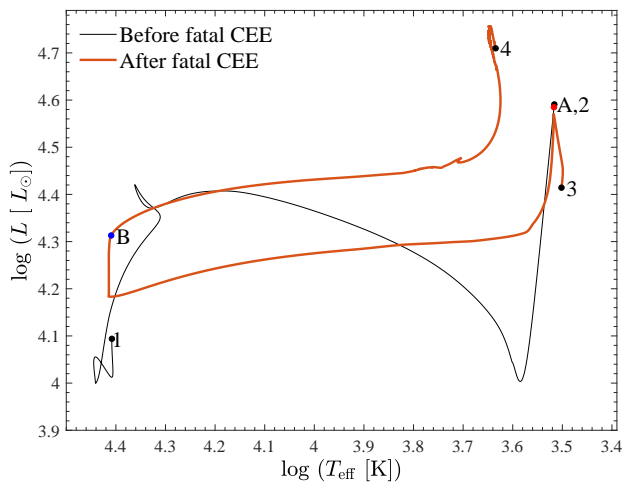}}
\caption{ The evolution on the HRD of the $M_{\rm ZAMS} = 12 M_\odot$ star before and after the CEE. Lines and marks are as in Fig. \ref{fig:HRAdd16}.}
\label{fig:HRAdd12}
\end{figure}

We repeat our treatment as described in the previous sections for a $12M_\odot$ primary star with rotation. In  Fig. \ref{fig:HRAdd12} we present the $M_{\rm ZAMS} = 12 M_\odot$ star on the HRD, similar to Fig. \ref{fig:HRAdd16}. We find the hydrogen mass at core collapse to be $M_{\rm H,exp,rot}=0.082 M_\odot$. {{{{ In both models ($16 M_\odot$ and $12 M_\odot$) the final luminosity of $\approx 10^5 L_\odot$ is reasonable for the pre-SN stage of stars whose mass is mostly their helium core and have thin hydrogen envelopes, e.g., figure 3 of \cite{Gilkisetal2019}. }}}}

\section{Discussion and summary}
\label{sec:summary}
 
We studied the fatal CEE channel for the formation of SN~IIb progenitors. In this channel a low mass, $M_2 \approx 0.5-1 M_\odot$, main sequence (MS) secondary star inspirals inside the giant envelope of the primary star and removes most of the giant envelope before it merges with the giant core. The core tidally destroys the MS companion. The gas of the destroyed secondary star forms a new giant envelope that contains little mass of hydrogen at core collapse (explosion).  

\cite{Nomotoetal1995} and \cite{Youngetal2006} already considered the fatal common envelope evolution (CEE) channel to form SNe~IIb. Our addition is the consideration of the formation of a new envelope from the destroyed secondary star. We followed the evolution of the merger product until core collapse (explosion). Our study is complementary to these earlier studies, and strengthens the merit of this scenario.

We studied in detail the case of a primary star with a ZAMS mass of $M_{\rm ZAMS} = 16 M_\odot$. In Fig. \ref{fig:initial16} we present some stellar properties after the rapid expansion of the star to become a giant. We showed (Fig. \ref{fig:binding}) that a secondary star can remove most of the envelope if $\alpha_{\rm CE} M_2 \ga 0.2 M_\odot$, where $\alpha_{\rm CE}$ is the common envelope parameter. We then removed the mass above the mass coordinate $M=4.8 M_\odot$ which corresponds to a radius of $r\simeq 1.5 R_\odot$ (Fig. \ref{fig:initial16}) at the beginning of the CEE. At that radius the core tidally destroys a low mass MS companion. We mimicked the merger process by adding the companion mass of $M_2=0.5 M_\odot$ to the almost bare core of the primary star. Due to numerical difficulties, we added the companion mass over a time period {{{{ of $50 \yr$, which although much longer than the dynamical time scale is much shorter than the thermal time scale of the new envelope. }}}}
 
We present the structure of the merger product at the end of the mass addition phase in Fig. \ref{fig:AfterAdd16}, and its evolution in Fig. \ref{fig:RadiusEvolution16}.  In Figs. \ref{fig:HRAdd16} and \ref{fig:HRAdd12} we present the evolution of the star before the CEE and after the merger in the HRD, for the $M_{\rm ZAMS} = 16 M_\odot$ and $M_{\rm ZAMS} = 12 M_\odot$ models, respectively. 

For the $M_{\rm ZAMS} = 16 M_\odot$ model we found that the hydrogen mass at core collapse is $M_{\rm H,exp,rot}=0.085 M_\odot$ for the evolution with rotation, and $M_{\rm H,exp}=0.024 M_\odot$ in the case without rotation. For the $M_{\rm ZAMS} = 12 M_\odot$ model we run the case with rotation only, and found $M_{\rm H,exp,rot}=0.082 M_\odot$. 

Despite some uncertainties in our calculations, we conclude that the fatal CEE scenario can account for some SNe~IIb.

The main uncertainties in our calculations are as follows.  

(1) \textit{The outcome of the CEE.} The first uncertainty concerns the envelope ejection before and during the CEE. We incorporated the CEE uncertainties in the commonly used $\alpha_{\rm CE}$ parameter. However, the CEE is not fully solved, and recent studies have raised questions and found difficulties in removing the common envelope and in explaining the final orbital separation (e.g., \citealt{GlanzPerets2018, Ivanova2018, MacLeodetal2018b, Sokeretal2018, Reichardtetal2019, IaconiDeMarco2019}, for a sample of papers just from the last year). The outcome depends on several factors, such as on the evolution just before the CEE (e.g., \citealt{BearSoker2010, IaconiDeMarco2019}), and on the question of whether the companion launches jets that help in removing the common envelope (e.g., \citealt{Shiberetal2019}).

(2) \textit{The merger process.} The second uncertainty involves the core-companion merger process, as we do not know how much mass might be lost during the process. If a large fraction of the mass is lost, e.g., in jets and disc outflow from the merging core-companion system, then we can allow for a somewhat more massive companion. But we cannot allow for a too massive companion, as such a companion can remove the entire envelope before it merges with the core.
{{{{ Let us elaborate more on the mass loss during the merger. In the merger process the core tidally destroys the MS companion to form a torus (thick disc). This merger processes releases a large amount of gravitational energy that ends mainly in two ways (as the gas is optically thick). At very early times the torus might launch jets (or disc winds). Part of this mass, if it does not collide with a slower gas, leaves the system. At the same time, the large amount of gravitational energy and the nuclear energy from the nuclear burning on the core inflates the torus to an extended envelope. Only three-dimensional hydrodynamical simulations can determine the amount of mass that the merger process directly ejects. At this stage we only estimate that it is at most a fraction of $0.1$--$0.2$ of the mass of the secondary star (as jets usually carry). }}}}

(3) \textit{Wind mass-loss rate.} The third uncertainty involves the mass-loss rate of the merger product (section \ref{sec:Numerics}). The uncertainty in the mass loss rate is common also to single star evolution. 

These uncertainties do not preclude the fatal CEE scenario, but they  make it difficult to estimate the exact ranges of secondary masses and initial orbital separations that lead binary systems to form SNe~IIb through the fatal CEE scenario. This in turn implies that we have a hard time in estimating the fraction of SNe~IIb that comes from the fatal CEE scenario. \cite{Soker2019FatalCEE} estimates that the fatal CEE scenario accounts for $\approx 10-30 \%$ of SNe~IIb, or $1-3 \%$ of all CCSNe (as SNe~IIb amount to $\approx 10 \%$ of all CCSNe, e.g., \citealt{Shivversetal2017}). Other channels, like single-star evolution, the RLOF scenario (e.g., \citealt{Sravanetal2019,Gilkisetal2019}), the grazing envelope evolution scenario \citep{Soker2017}, and a scenario of a CEE where the companion survives, account for the rest.

\section*{Acknowledgments}

{{{{ We thank an anonymous referee for very helpful comments. }}}}
This research was supported by the Israel Science Foundation. AG gratefully acknowledges the generous support of the Blavatnik Family Foundation.

\label{lastpage}

\begin{thebibliography}{}

\bibitem[Aldering et al.(1994)]{Alderingetal1994} Aldering, G., Humphreys, R.~M., \& Richmond, M.\ 1994, \aj, 107, 662

\bibitem[Bear \& Soker(2010)]{BearSoker2010} Bear, E., \& Soker, N.\ 2010, \na, 15, 483

\bibitem[Benvenuto et al.(2013)]{Benvenutoetal2013} Benvenuto, O.~G., Bersten, M.~C., \& Nomoto, K.\ 2013, \apj, 762, 74


\bibitem[Chevalier(2012)]{Chevalier2012} Chevalier, R.~A.\ 2012, \apjl, 752, L2

\bibitem[Chevalier, \& Soderberg(2010)]{ChevalierSoderberg 2010} Chevalier, R.~A., \& Soderberg, A.~M.\ 2010, \apjl, 711, L40

\bibitem[Chevalier \& Soker(1989)]{ChevalierSoker1989} Chevalier, R.~A., \& Soker, N.\ 1989, \apj, 341, 867 

\bibitem[Claeys et al.(2011)]{Claeys2011} Claeys, J.~S.~W., de Mink, S.~E., Pols, O.~R., Eldridge, J.~J., \& Baes, M.\ 2011, \aap, 528, A131

\bibitem[{de Jager} et al.(1988)]{deJager1988} {de Jager}, C., Nieuwenhuijzen, H., \& {van der Hucht} K.~A.\ 1988, \aaps, 72, 259

\bibitem[Fox et al.(2014)]{Foxetal2014} Fox, O.~D., Azalee Bostroem, K., Van Dyk, S.~D., et al.\ 2014, \apj, 790, 17

\bibitem[Georgy et al.(2013)]{Georgyetal2013} Georgy, C., Ekstr{\"o}m, S., Eggenberger, P., et al.\ 2013, \aap, 558, A103

\bibitem[Gilkis et al.(2019)]{Gilkisetal2019} Gilkis, A., Vink, J.~.S., Eldridge, J.~J., \& Tout, C.~A.\ 2019, \mnras, 486, 4451

\bibitem[Glanz \& Perets(2018)]{GlanzPerets2018} Glanz, H., \& Perets, H.~B.\ 2018, \mnras, 478, L12

\bibitem[Gorlova et al.(2012)]{Gorlovaetal2012} Gorlova, N., Van Winckel, H., Gielen, C., et al.\ 2012, \aap, 542, A27

\bibitem[Gorlova et al.(2015)]{Gorlovaetal2015} Gorlova, N., Van Winckel, H., Ikonnikova, N.~P., Burlak, M.~A., Komissarova, G.~V., Jorissen, A., Gielen, C., Debosscher, J., \& Degroote, P. 2015\, \mnras, 451, 2462

\bibitem[Graur et al.(2017)]{Grauretal2017} Graur, O., Bianco, F.~B., Modjaz, M., Shivvers, I., Filippenko, A.~V., Li, W., Smith, N.\ 2017, \apj, 837, 121

\bibitem[Harpaz \& Soker(1994)]{HarpazSoker1994} Harpaz, A., \& Soker, N.\ 1994, \mnras, 270, 734 

\bibitem[Iaconi \& De Marco(2019)]{IaconiDeMarco2019} Iaconi, R., \& De Marco, O.\ 2019, arXiv:1902.02039

\bibitem[Ivanova \& Podsiadlowski(2002)]{IvanovaPodsiadlowski2002} Ivanova, N., \& Podsiadlowski, P.\ 2002, \apss, 281, 191 

\bibitem[Ilkov \& Soker(2012)]{IlkovSoker2012} Ilkov, M., \& Soker, N.\ 2012, \mnras, 419, 1695

\bibitem[Ivanova(2018)]{Ivanova2018} Ivanova, N.\ 2018, \apjl, 858, L24

\bibitem[Joss et al.(1988)]{Jossetal1988} Joss P.~C., Podsiadlowski P., Hsu J.~J.~L., Rappaport S., 1988, Natur, 331, 237
 
\bibitem[Kastner et al.(2010)]{Kastneretal2010} Kastner, J.~H., Buchanan, C., Sahai, R., Forrest, W.~J., \& Sargent, B.~A.\ 2010, \aj, 139, 1993

\bibitem[Kerzendorf et al.(2019)]{Kerzendorfetal2019} Kerzendorf, W.~E., Do, T., {de Mink}, S.~E.,  {G{\"o}tberg}, Y., Milisavljevic, D., Zapartas, E., Renzo, M., Justham, S., Podsiadlowski, P., Fesen, R.~A.\ 2019, \aap, 623, 34

\bibitem[Kilpatrick et al.(2017)]{Kilpatricketal2017} Kilpatrick, C.~D., et al.\ 2017, \mnras, 465, 4650

\bibitem[Kochanek(2018)]{Kochanek2018} Kochanek, C.~S.\ 2018, \mnras, 473, 1633

\bibitem[Matheson et al.(2000)]{Mathesonetal2000} Matheson, T., Filippenko, A.~V., Ho, L.~C., Barth, A.~J., \& Leonard, D.~C.\ 2000, \aj, 120, 1499

\bibitem[MacLeod et al.(2018)]{MacLeodetal2018b} MacLeod, M., Ostriker, E.~C., \& Stone, J.~M.\ 2018, \apj, 868, 136 

\bibitem[Menon \& Heger(2017)]{MenonHeger2017} Menon, A., \& Heger, A.\ 2017, \mnras, 469, 4649

\bibitem[Menon et al.(2019)]{Menonetal2019} Menon, A., Utrobin, V., \& Heger, A.\ 2019, \mnras, 482, 438

\bibitem[Meynet et al.(2015)]{Meynetetal2015} Meynet, G., Chomienne, V., Ekstr{\"o}m, S., et al.\ 2015, \aap, 575, A60

\bibitem[Naiman et al.(2019)]{Naimanetal2019} Naiman, B.~V., Sabach, E., Gilkis, A., \& Soker, N.\ 2019, arXiv:1909.04583 

\bibitem[Nakaoka et al.(2019)]{Nakaokaetal2019} Nakaoka, T., Moriya, T.~J., Tanaka, M. et al.\ 2019, \apj, 875, 76
  
\bibitem[Nomoto et al.(1995)]{Nomotoetal1995} Nomoto, K.~I.;,Iwamoto, K., \& Suzuki, T.\ 1995, Physics Reports, 256, 173

\bibitem[Nomoto et al.(1993)]{Nomotoetal1993} Nomoto K., Suzuki T., Shigeyama T., Kumagai S., Yamaoka H., Saio H., 1993, Natur, 364, 507

\bibitem[Ouchi \& Maeda(2017)]{OuchiMaeda2017} {{{ Ouchi, R., \& Maeda, K.\ 2017, \apj, 840, 90}}}

\bibitem[Paxton et al.(2011)]{Paxtonetal2011} Paxton, B., Bildsten, L., Dotter, A., Herwig, F., Lesaffre, P., \& Timmes, F.\ 2011, \apjs, 192, 3

\bibitem[Paxton et al.(2013)]{Paxtonetal2013} Paxton, B., Cantiello, M., Arras, P., et al.\ 2013, \apjs, 208, 4

\bibitem[Paxton et al.(2015)]{Paxtonetal2015} Paxton, B., Marchant, P., Schwab, J., et al.\ 2015, \apjs, 220, 15

\bibitem[Paxton et al.(2018)]{Paxtonetal2018} Paxton, B., Schwab, J., Bauer, E.~B., et al.\ 2018, \apjs, 234, 34

\bibitem[Podsiadlowski et al.(1993)]{Podsiadlowskietal1993} Podsiadlowski, P., Hsu, J.~J.~L., Joss, P.~C., \& Ross, R.~R.\ 1993, \nat, 364, 509

\bibitem[Podsiadlowski et al.(1992)]{Podsiadlowskietal1992} Podsiadlowski, P., Joss, P.~C., \& Hsu, J.~J.~L.\ 1992, \apj, 391, 246

\bibitem[Podsiadlowski et al.(1990)]{Podsiadlowskietal1990} Podsiadlowski, P., Joss, P.~C., \& Rappaport, S.\ 1990, \aap, 227, L9

\bibitem[Reichardt et al.(2019)]{Reichardtetal2019} Reichardt, T.~A., De Marco, O., Iaconi, R., Tout, C.~A., \& Price, D.~J.\ 2019, \mnras, 484, 631

\bibitem[Rest et al.(2011)]{Restetal2011} Rest, A., Foley, R.~J., Sinnott, B., et al.\ 2011, \apj, 732, 3 

\bibitem[Segev et al.(2019)]{Segevetal2019} Segev, R., Sabach, E., \& Soker, N.\ 2019, preprint

\bibitem[Shiber et al.(2019)]{Shiberetal2019} Shiber, S., Iaconi, R., De Marco, O., \& Soker, N.\ 2019, arXiv:1902.03931

\bibitem[Shivvers et al.(2017)]{Shivversetal2017} Shivvers, I., Modjaz, M., Zheng, W., et al.\ 2017, \pasp, 129, 054201

\bibitem[Siess \& Livio(1999)]{SiessLivio1999} Siess, L., \& Livio, M.\ 1999, \mnras, 308, 1133 

\bibitem[Smith et al.(2011)]{Smithetal2011} Smith, N., Li, W., Filippenko, A.~V., \& Chornock, R.\ 2011, \mnras, 412, 1522

\bibitem[Soker(2017)]{Soker2017}  Soker, N.\ 2017, \mnras, 470, L102

\bibitem[Soker(2019)]{Soker2019FatalCEE} Soker, N.\ 2019, Sci. China Phys. Mech. Astron., 62, 9501

\bibitem[Soker \& Gilkis(2018)]{SokerGilkis2018} Soker, N., \& Gilkis, A.\ 2018, \mnras, 475, 1198
  
\bibitem[Soker et al.(2018)]{Sokeretal2018} Soker N., Grichener A., \& Sabach E.,\ 2018, \apj, 863, L14

\bibitem[Sravan(2016)]{Sravan2016} Sravan, N.\ 2016, 41st COSPAR Scientific Assembly, 41,

\bibitem[Sravan et al.(2019)]{Sravanetal2019} Sravan, N., Marchant, P., \& Kalogera, V.\ 2019, arXiv:1808.07580

\bibitem[Stancliffe \& Eldridge(2009)]{StancliffeEldridge2009} Stancliffe, R.~J., \& Eldridge, J.~J.\ 2009, \mnras, 396, 1699

\bibitem[Thomas et al.(2013)]{Thomasetal2013} Thomas, J.~D., Witt, A.~N., Aufdenberg, J.~P., Bjorkman, J. E., Dahlstrom, J. A., Hobbs, L. M., \& York, D. G.\ 2013, \mnras, 430, 1230

\bibitem[Urushibata et al.(2018)]{Urushibataetal2018} Urushibata, T., Takahashi, K., Umeda, H., \& Yoshida, T.\ 2018, \mnras, 473, L101 

\bibitem[Van Winckel(2017a)]{VanWinckel2017} Van Winckel, H.\ 2017a, Astronomical Society of the Pacific Conference Series, 508, 197

\bibitem[Van Winckel(2017b)]{VanWinckel2017b} {{{ Van Winckel, H.\ 2017b, in Planetary Nebulae: Multi-Wavelength Probes of Stellar and Galactic Evolution, Proceedings IAU Symposium No. 323,  eds. X. Liu, L. Stanghellini, and A. Karakas A.C., in press   }}}

\bibitem[Vink et al.(2001)]{Vink2001} {Vink}, J.~S., {de Koter}, A., \& Lamers, H.~J.~G.~L.~M.\ 2001, \aap, 369, 574

\bibitem[Witt et al.(2009)]{Wittetal2009} Witt, A.~N., Vijh, U.~P., Hobbs, L.~M., Aufdenberg, J. P., Thorburn, J. A., \& York, D. G.\ 2009, \apj, 693, 1946

\bibitem[Woosley et al.(1994)]{Woosleyetal1994} Woosley, S.~E., Eastman, R.~G., Weaver, T.~A., \& Pinto, P.~A.\ 1994, \apj, 429, 300

\bibitem[Yoon et al.(2017)]{Yoonetal2017} Yoon, S.-C., Dessart, L., \& Clocchiatti, A.\ 2017, \apj, 840, 10

\bibitem[Young et al.(2006)]{Youngetal2006} Young, P.~A., Fryer, C.~L., Hungerford, A., et al.\ 2006, \apj, 640, 891

\end{thebibliography}
\end{document}